\documentclass[9pt,twocolumn,twoside,fleqn]{osajnl}
\journal{ol} 
\setboolean{shortarticle}{true}

\title{Sum rules for energy deposition eigenchannels in scattering systems}

\author[1,*]{Alexey Yamilov}
\author[2]{Nicholas Bender}
\author[3]{Hui Cao}

\affil[1]{Physics Department, Missouri University of Science \& Technology, Rolla, Missouri, USA}
\affil[2]{School of Applied and Engineering Physics, Cornell University, Ithaca, New York, USA}
\affil[3]{Department of Applied Physics, Yale University, New Haven, Connecticut, USA}
\affil[*]{Corresponding author: yamilov@mst.edu}

\begin{abstract}
	In a random-scattering system, the deposition matrix maps the incident wavefront to the internal field distribution across a target volume. The corresponding eigenchannels have been used to enhance the wave energy delivered to the target. Here we find the sum rules for the eigenvalues and eigenchannels of the deposition matrix in any system geometry: including two and three-dimensional scattering systems, as well as narrow waveguides and wide slabs.	We derive a number of constraints on the eigenchannel intensity distributions inside the system as well as the corresponding eigenvalues. Our results are general and applicable to random systems of arbitrary scattering strength as well as different types of waves including electromagnetic waves, acoustic waves, and matter waves.
\end{abstract}
\setboolean{displaycopyright}{true}

\begin{document}
	\maketitle
	
	
	
	\section*{Introduction}\label{sec:intro}
	
	Wavefront shaping opened a new frontier for coherent manipulation of wave propagation in complex media~\cite{2012_Mosk_SLM_review, 2015_Yu_Wavefront_Shaping_Review, kubby2019wavefront,  2021_Gigan_Roadmap}. The core idea is rooted in the determinism of coherent wave propagation in static, linear scattering media~\cite{2017_Rotter_Gigan_review}. Although the incident wavefront can be optimized via an iterative optimization procedure~\cite{2015_Vellekoop_review}, solving the eigenvalue problem of a linear operator (matrix) is a more predictive approach~\cite{2010_Popoff_Shaping_PRL}. Recently, a number of matrices have been introduced in order to manipulate quantities such as transmittance, reflectance, dwell-time, spatial distribution etc, see Refs.~(\cite{2015_Yang_Wavefront_Shaping_Guidestar, 2017_Rotter_Gigan_review, yoon2020deep, 2021_Gigan_Roadmap}) for review. Control and optimization of wave energy inside a scattering medium requires an access to the internal field distribution~\cite{2014_Cheng_Wavefront_Shaping_Energy_Deposition}. To find the ultimate limit of energy delivery to a target buried deep inside a diffusive medium, we recently introduced a deposition matrix that relates the incident wavefront to the internal field distribution across the target~\cite{2022_Bender_Deposition_eigenchannels}. The maximal eigenvalue of the matrix gives the largest possible energy enhancement, while the corresponding eigenvector gives the optimal incident wavefront.  However, the properties of deposition eigenvalues and eigenchannels remain essentially unknown.
	
	In this work, we theoretically obtain a series of sum rules satisfied by the deposition eigenchannels and their eigenvalues. In Ref.~\cite{2022_Bender_Deposition_eigenchannels} a deposition eigenchannel is decomposed by the transmission eigenchannels to illustrate the incoherent (i.e. intensity summation) and coherent (interference) contributions. Here, we demonstrate that these contributions also obey rigorous sum rules. These relationships not only provide physical insights into the deposition eigenchannels, but also play an important role in utilizing such channels for the most efficient energy delivery in experiments.
	
	\begin{figure}[htbp]
		\vskip -0.2cm
		\centering
		\includegraphics[width=3in]{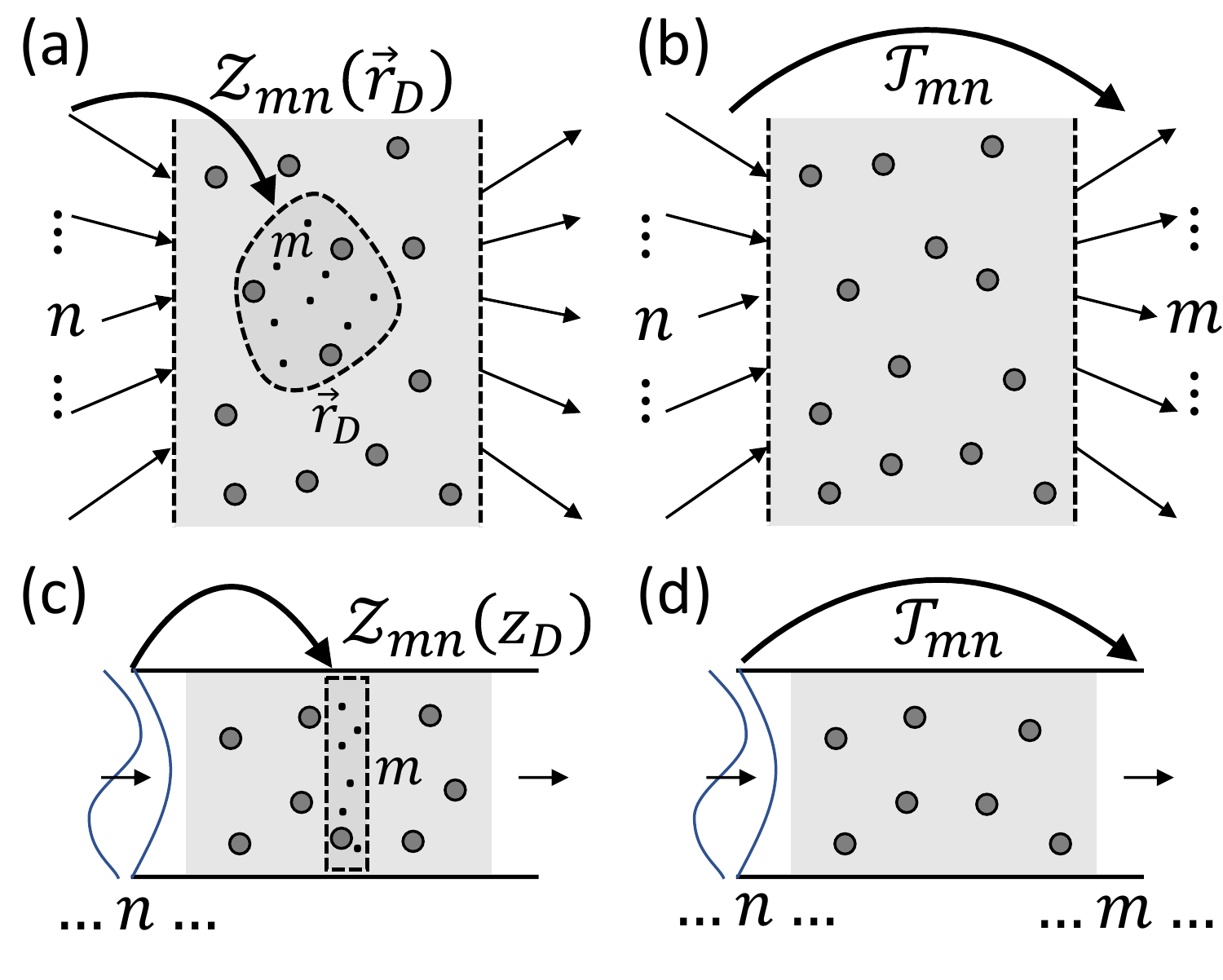}
		\vskip -0.2cm
		\caption{Schematic depiction of the slab (a,b) and the waveguide scattering-medium geometries. The deposition matrix relates the input to the internal fields inside a scattering medium, panels (a,c). The transmission matrix $\mathcal{T}_{mn}$ relates the input and output degrees of freedom, panels (b,d).}
		\label{fig:schematics}
	\end{figure}
	
	\section*{Deposition matrix}\label{sec:definitions}
	  
	Fig.~\ref{fig:schematics} illustrates two geometries we consider for energy delivery into a linear scattering system: (a,b) wide slab with open (leaky) boundary, (c,d) narrow waveguide with closed (reflecting) sidewall. The deposition matrix (DM) $\mathcal{Z}$ is introduced for a target region of arbitrary size, shape, and depth~\cite{2022_Bender_Deposition_eigenchannels}. It relates an orthonormal set of input waves to the corresponding spatial field distributions within the target region. As shown in  Fig.~\ref{fig:schematics}(a) of a wide slab, the orthogonal input modes have distinct wavevectors. While in Fig.~\ref{fig:schematics}(c) the input waves are a complete set of waveguide modes. The total number of input modes is $N$. With a unit flux of incident light in the $n$-th mode, the complex field distribution throughout the scattering system is $E^{(0)}_n(\vec{r})$. We sample the fields uniformly across a target region of volume $\mathcal{V}$ centered at $\vec{r}_D$, see Figs.~\ref{fig:schematics}(a). The field at the $m$-th sampling point $\vec{r}_m$ is $E^{(0)}_n(\vec{r}_m;\vec{r}_D)$, where $m=\{1,...,M\}$. The volume $\mathcal{V}/{M}$ covered by each sampling point is much smaller than $\lambda^3$, where $\lambda$ is the wavelength. The deposition matrix of dimension $M \times N$ is defined as
	\begin{equation} 
	\mathcal{Z}_{mn}(\vec{r}_D) \equiv \left[\epsilon(\vec{r}_m) \, \frac{\mathcal{V}}{M} \right]^{1/2} \, E^{(0)}_n(\vec{r}_m;\vec{r}_D) \, ,
	\label{eq:Z_definition}
	\end{equation}	 
	where $\epsilon(\vec{r})$ is the spatially-varying dielectric constant. 	
	
	The deposition eigenchannels are obtained from the singular value decomposition (SVD), $\mathcal{Z}_{mn}(\vec{r}_D)=\sum_{\alpha=1}^{N}U_{m\alpha}^{(D)}(\vec{r}_D) \, \zeta_\alpha^{1/2}(\vec{r}_D) \, \left[V_{\alpha n}^{(D)}(\vec{r}_D)\right]^*$. The incident wavefront of the $\alpha$-th eigenchannel is $E_{\alpha}^{(D)}(\vec{r};\vec{r}_D) = \sum_{n=1}^{N} E^{(0)}_{n}(\vec{r})\, V_{n\alpha}^{(D)}(\vec{r}_D)$.
	
	$\zeta_\alpha(\vec{r}_D)$ and $V_{n\alpha}^{(D)}(\vec{r}_D)$ are the eigenvalue and eigenvector of matrix $\mathcal{Z}^{\dagger}(\vec{r}_D)\mathcal{Z}(\vec{r}_D)$, respectively. Using the unitarity of matrices $U^{(D)}(\vec{r}_D)$ and $V^{(D)}(\vec{r}_D)$ as well as the definition of $\mathcal{Z}(\vec{r}_D)$ in Eq.~(\ref{eq:Z_definition}), we get an explicit relationship for the eigenvalue~\cite{SM} 
	\begin{equation}
	\zeta_\alpha(\vec{r}_D)=
	\frac{\mathcal{V}}{M}\,\sum_{m=1}^{M} \epsilon(\vec{r}_m)
	\left|E_\alpha^{(D)}(\vec{r}_m;\vec{r}_D)\right|^2. 
	\label{eq:zeta_definition}
	\end{equation}
	This relationship reveals that the deposition eigenvalue is equal to the total energy inside the target region with a coherent excitation of $E_{\alpha}^{(D)}(\vec{r};\vec{r}_D)$. 
	
	For comparison, Fig.~\ref{fig:schematics}(b,d) shows the transmission matrix $\mathcal{T}$ that maps the incident fields to the transmitted fields. When a monochromatic light with a unit flux in the $n$-th mode is incident to the scattering system, $\left|\mathcal{T}_{mn}\right|^2$ is the amount of flux carried away by the $m$-th outgoing mode in transmission. The transmission eigenchannels are obtained from the SVD $\mathcal{T}_{mn}=\sum_{\alpha=1}^{N} U_{m\alpha}^{(T)} \cdot \tau_{\alpha}^{1/2} \cdot \left[V_{\alpha n}^{(T)}\right]^{*}$. The input and output wavefronts of the $\alpha$-th transmission eigenchannel are given by $V_{\alpha n}^{(T)}$ and $U_{m\alpha}^{(T)}$, and the transmittance by eigenvalue $\tau_{\alpha}$.
	
	\section*{Sum rules for deposited energy}\label{sec:I_sum}
	
	Singular value decomposition is a linear transformation of the input basis that yields an orthonormal set defined by the unitary matrix $V^{(D)}(\vec{r}_D)$ or $V^{(T)}$ for the deposition and transmission, respectively. This orthonormal property of the input eigenvectors $V_{n\alpha}^{(D)}(\vec{r}_D)$ and $V_{n\alpha}^{(T)}$ has a profound impact on the distribution of the wave intensity {\it inside} the scattering medium. Summing the intensity of all eigenchannels at any given point $\vec{r}$ gives~\cite{SM}
	\begin{equation}
	\sum_{\alpha=1}^{N}\left| E_{\alpha}^{(D)}(\vec{r};\vec{r}_D) \right|^2
	=\sum_{n=1}^{N}\left| E_\alpha^{(T)}(\vec{r}) \right|^2
	=\sum_{n=1}^{N}\left| E^{(0)}_n(\vec{r}) \right|^2.
	\label{eq:DE_summation_rule}
	\end{equation}
	The sum of intensities by subsequently exciting the system with individual deposition or transmission eigenchannels is equal to that with any orthogonal set of input modes. The above relationships lead to two remarkable properties. First, the physical quantity being preserved by SVD transformation is the intensity, and not the field. Because Eq.~(\ref{eq:DE_summation_rule}) holds at every position, and thus can be multiplied by $\epsilon(\vec{r}) $ from both sides, it can be interpreted as {\it point-wise} (i.e. local) conservation of energy. Secondly, the above relationship is not statistical -- it does not involve any statistical averaging over an ensemble of disorder realizations, instead, it holds for every realization.
	
	\begin{figure}[htbp]
		\centering
		\includegraphics[width=\linewidth]{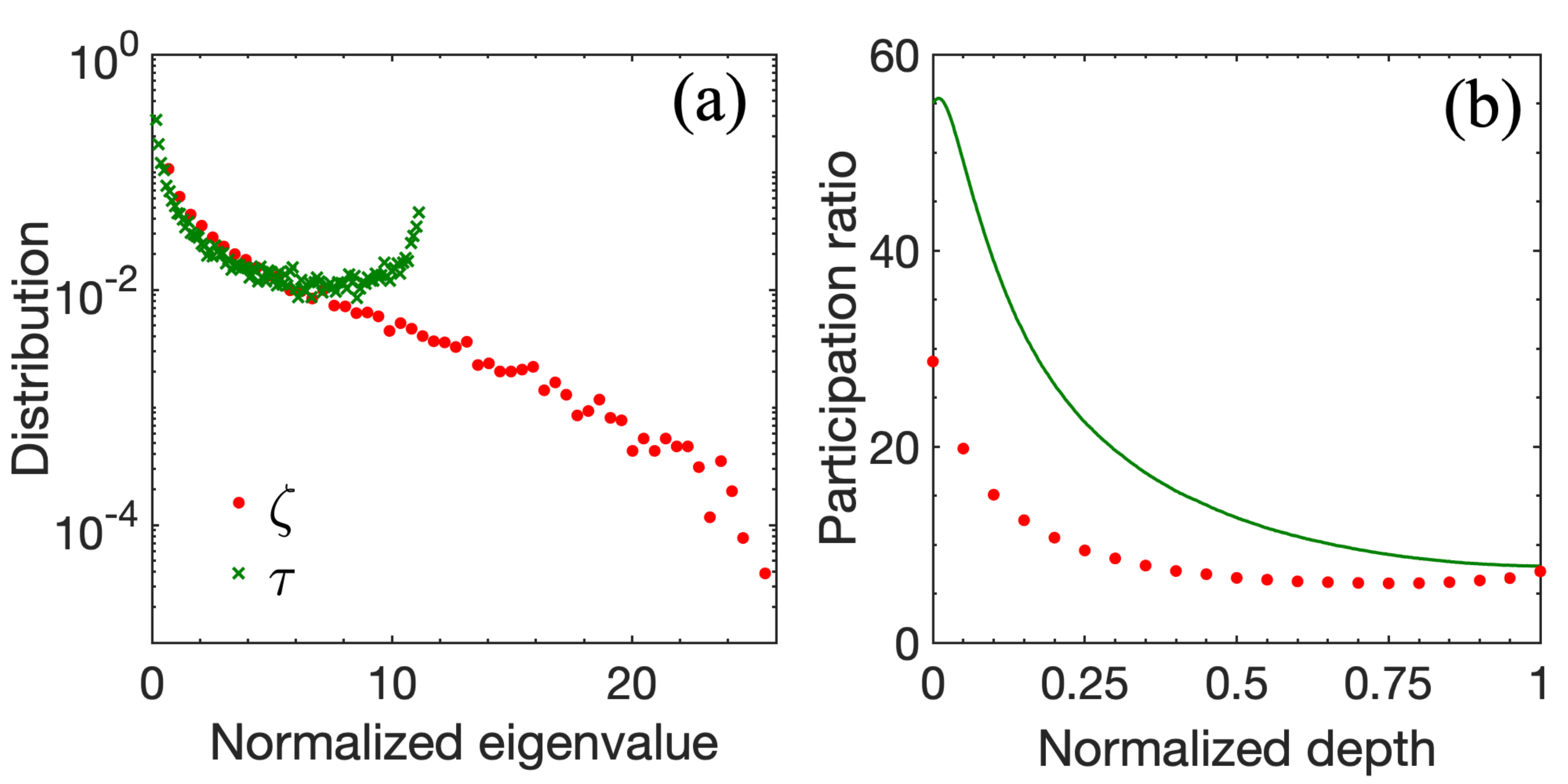}
		\vskip -0.2cm
		\caption{(a) Numerically computed probability density of the deposition (red circle) and transmission (green cross)  eigenvalues for a slice target at the depth $z_D/L=1/2$ in a 2D disordered waveguide~\cite{SM}. (b) Intensity participation ratio for deposition and transmission eigenchannels at different depths, see text.}
		\label{fig:eigenvalues}
	\end{figure}
	
\begin{figure}[htbp]
		\centering
		\includegraphics[width=\linewidth]{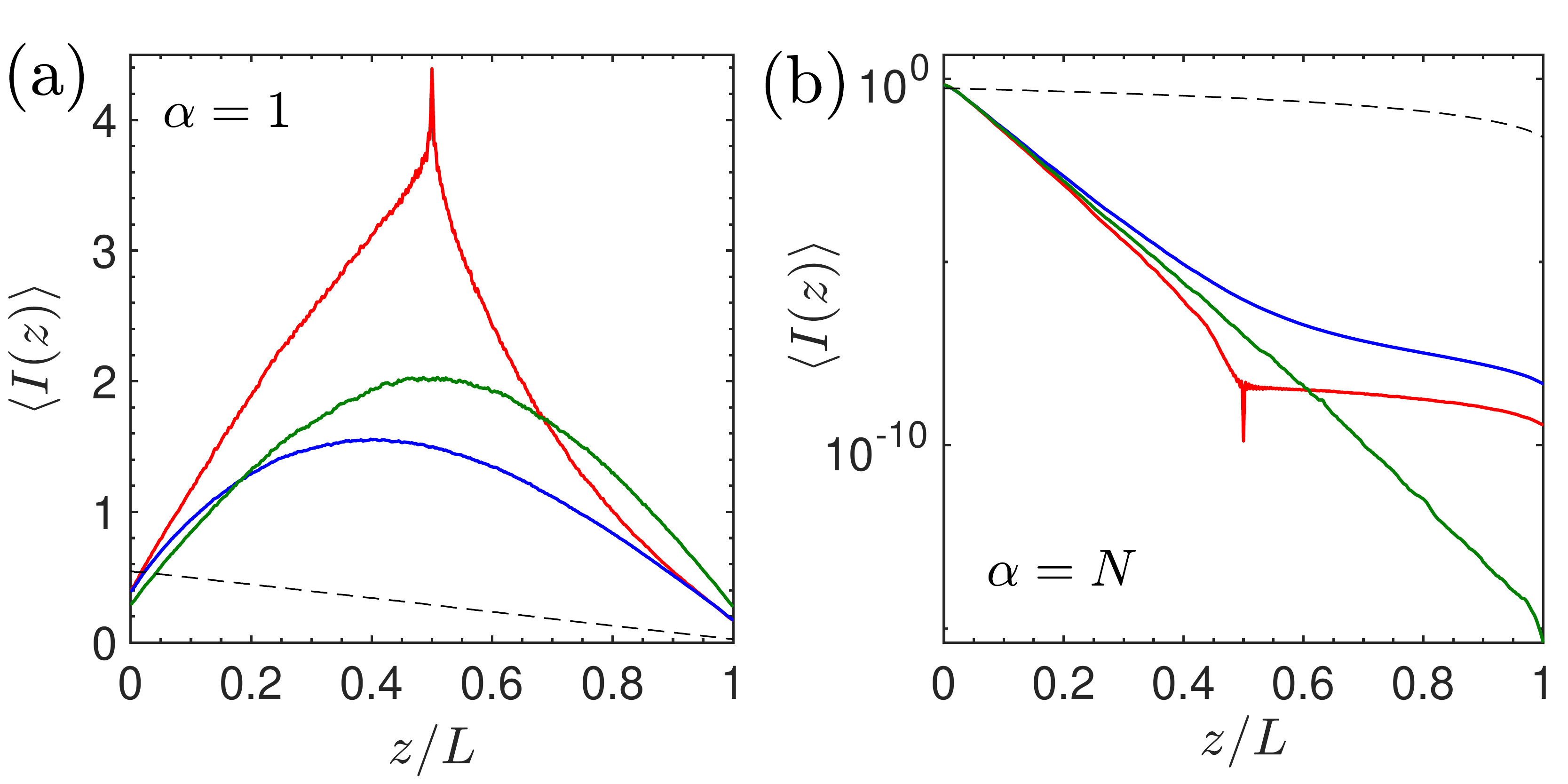}
		\vskip -0.1cm
		\caption{Deposition eigenchannels (red) in disordered waveguide geometry~\cite{SM} with $\alpha=1,N$ for $z_D/L=1/2$ and their incoherent components (blue) are compared to the corresponding TEs (green). Random input intensity profile (dashed line) is shown for reference. Around the target region, the enhancement/suppression of the local intensity of the DEs above/below that of the maximum/minimum transmission eigenchannels is observed. The incoherent part, on the other hand, is below/above the maximum/minimum transmission eigenchannels.}
		\label{fig:eigenchannels}
	\end{figure}

	The sum rule in Eq.~(\ref{eq:DE_summation_rule}) can be exploited to estimate the intensity enhancement when combined with a known probability density function (PDF) of the corresponding eigenvalues. To illustrate this point, lets consider transmission eigenchannels of a diffusive system and use a numerical simulation of 2D disordered waveguide in Fig.~\ref{fig:schematics}(d). The PDF of the transmission eigenvalues in this case is the celebrated bimodal distribution~\cite{1984_Dorokhov}, predicting that only $g\ll N$ eigenchannels have $\tau_\alpha\sim 1$ while the rest have $\tau_\alpha\sim 0$, c.f. Fig.~\ref{fig:eigenvalues}a, where $g$ is the dimensionless conductance. In contrast, for such system, the PDF of deposition eigenvalues has been predicted in Ref.~\cite{2022_Bender_Deposition_eigenchannels} to exhibit a long tail toward the largest value of $\zeta_{max}(\vec{r}_D)/\langle \zeta(\vec{r}_D)\rangle$, see Fig.~\ref{fig:eigenvalues}a. Consequently, fewer deposition eigenchannels with very large $\zeta(\vec{r})$ provide a higher enhancement at the target region, compared to the high-transmission eigenchannels. This effect can be quantified using the intensity participation ratio ${\cal P}^{(D)}(z_D)=\left(\sum_{\alpha=1}^N \left| E_{\alpha}^{(D)}(z_D;z_D) \right|^2 \right)^2 / \left(\sum_{\alpha=1}^N \left| E_{\alpha}^{(D)}(z_D;z_D) \right|^4 \right)$ for deposition eigenchannels, and ${\cal P}^{(T)}(z)=\left(\sum_{\alpha=1}^N \left| E_{\alpha}^{(T)}(z) \right|^2 \right)^2 / \left(\sum_{\alpha=1}^N \left| E_{\alpha}^{(T)}(z) \right|^4 \right)$ for transmission eigenchannels~\cite{2015_Genack_Eigenchannels_Inside}. Fig.~\ref{fig:eigenchannels}b indeed shows a lower participation ratio ${\cal P}^{(D)}$ than ${\cal P}^{(T)}$, reflecting a smaller number of deposition eigenchannels with major contributions. This, in combination with the sum rule in Eq.~(\ref{eq:DE_summation_rule}), leads to a higher cross-section integrated intensity at the target depth $z_D$, see also Fig.~\ref{fig:eigenchannels}a. Furthermore, the extraordinarily large intensities of a few deposition eigenchannels, in turn, allow for the minimal intensity to be even below that of the lowest-transmission eigenchannel, c.f. Fig.~\ref{fig:eigenchannels}b. All of these conclusions are indeed supported by Ref.~\cite{2022_Bender_Deposition_eigenchannels}, demonstrating that a combination of the knowledge of the eigenvalue PDF and the constraints imposed by Eq.~(\ref{eq:DE_summation_rule}) has important implications for local energy density enhancement by the eigenchannels. 
	
	We also find the sum rule for the deposition eigenvalues. Summing over $\zeta_\alpha(\vec{r}_D)$ and recalling Eq.~(\ref{eq:DE_summation_rule}), we obtain
	\begin{eqnarray}
	\sum_{n=1}^{N}\zeta_\alpha(\vec{r}_D) & = &
	\frac{\mathcal{V}}{M}\,\sum_{m=1}^{M} \epsilon(\vec{r}_m)
	\left[
	\sum_{\alpha=1}^{N} \left|E_\alpha^{(D)}(\vec{r}_m;\vec{r}_D)\right|^2
	\right] \nonumber \\
	 & = & \frac{\mathcal{V}}{M}\,\sum_{m=1}^{M} \epsilon(\vec{r}_m)
	 \left[
	 \sum_{\alpha=1}^{N} \left|E_n^{(T)}(\vec{r}_m)\right|^2
	 \right] \nonumber \\
	 & = & \frac{\mathcal{V}}{M}\,\sum_{m=1}^{M} \epsilon(\vec{r}_m)
	 \left[
	 \sum_{n=1}^{N} \left|E_n^{(0)}(\vec{r}_m)\right|^2
	 \right]  
	\label{eq:zeta_summation_rule}
	\end{eqnarray}
	This equation shows that the sum of all deposition eigenvalues is equal to sum of local energy (sampled over $M$ points $\left\{\vec{r}_m\right\}$) excited by all deposition or transmission eigenchannels or any orthogonal set of input waves. 
	
	Therefore, the sum of all deposition eigenvalues gives the total energy within the target region excited by all input degrees of freedom. Moreover, the degree of control of the energy delivery via the deposition eigenchannels is determined by the PDF of deposition eigenvalue $P(\zeta)$. In diffusive systems, $P(\zeta)$ is amenable to a theoretical description within framework of filtered random matrix (FRM)  theory\cite{2013_Stone_Eigenvalues_with_Absorption,2022_Bender_Deposition_eigenchannels}. The maximal and minimal energies over the target are set by the range of $P(\zeta(\vec{r}_D))$.
	
	\section*{Interference of transmission eigenchannels}\label{sec:coherent_contribution}

	At first glance, the intensity sum rule in Eq.~(\ref{eq:DE_summation_rule}) might be misconstrued to mean that the intensity profile of a channel in one basis, e.g. the deposition eigenchannel, is a linear superposition of the intensity pattern in another basis, e.g. of the transmission eigenchannel. Such a conclusion is markedly wrong, as seen by comparing $\left| E^{(0)}_n(\vec{r})\right|^2$ and $\left| E_{\alpha}^{(D)}(\vec{r};\vec{r}_D) \right|^2$, and recalling that the former set is composed of the statistically identical intensity profiles. The fault of the above argument lies in neglecting the interference between channels inside the scattering system. Indeed, we can express
	\begin{equation}
	E_{\alpha}^{(D)}(\vec{r};\vec{r}_D)=\sum_{\beta=1}^{N}  E_\beta^{(T)}(\vec{r})  d_{\beta\alpha}(\vec{r}_D), 
	\label{eq:DE_decomposition}
	\end{equation}
	where the decomposition coefficient
	$d_{\beta\alpha}(\vec{r}_D) = \sum_{n^\prime=1}^{N}\left[V_{\beta n^\prime}^{(T)}\right]^{*} V_{n^\prime\alpha}^{(D)}(\vec{r}_D)$
	represents a projection of input vector of the $\alpha$-th deposition eigenchannel onto that of the $\beta$-th transmission eigenchannel. Using these decomposition coefficients, the intensity pattern of a deposition eigenchannel can be expressed in terms of two distinct terms:
	\begin{equation}
	\begin{split}
	\left|E_{\alpha}^{(D)}(\vec{r};\vec{r}_D)\right|^2  =   
	&\sum_{\beta=1}^N \left|E_{\beta}^{(T)}(\vec{r})\right|^2 |d_{\beta\alpha}|^2 \\
	+&\sum_{\beta \neq \beta^\prime}^N d_{\beta\alpha} \, d^*_{\alpha\beta^\prime} E_{\beta}^{(T)}(\vec{r}) \left[E_{\beta^\prime}^{(T)}(\vec{r})\right]^*. 
	\label{eq:DE_decomposition_coh_incoh}
	\end{split}
	\end{equation}
	The first term is an incoherent sum of TE intensity patterns, whereas the second term is the result of interference between different transmission eigenchannels inside the scattering medium. The numerical simulation results in Fig.~\ref{fig:eigenchannels}a,b illustrate that the interference contributions can be positive or negative in order to enhance or suppress the energy in the target. Their existence is essential to make the energy higher than the largest-transmission eigenchannel or lower than the smallest-transmission eigenchannel. 
	
	\begin{figure}[htbp]
		\vskip -0.4cm
		\centering
		\includegraphics[width=3.25in]{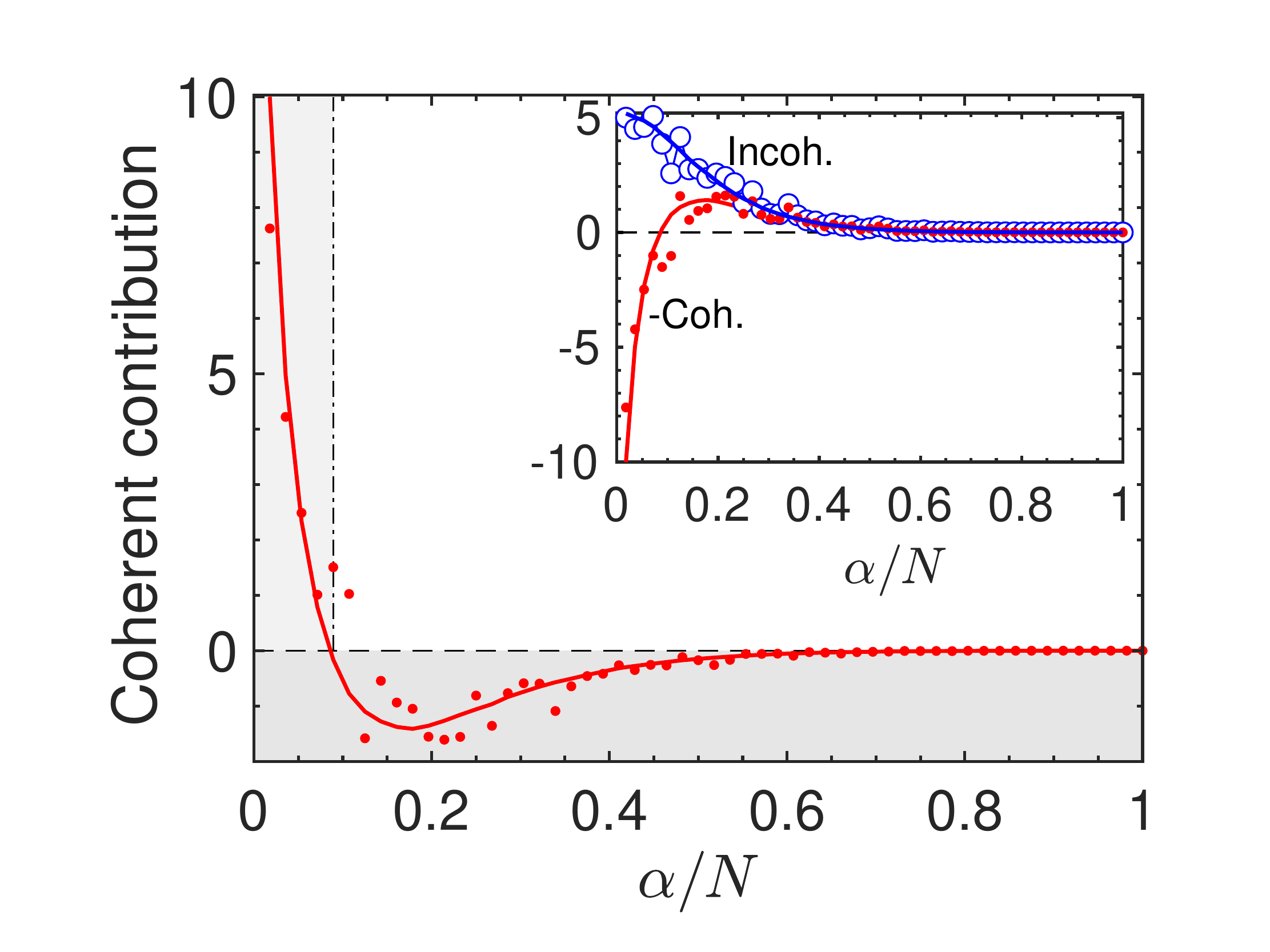}
		\vskip -0.3cm
		\caption{Coherent contribution in Eq.~(\ref{eq:DE_decomposition_coh_incoh}) for each deposition eigenchannel $\alpha$, computed at $z=z_D$ in numerical model~\cite{SM}, is shown in the main plot. A small number of large positive contributions is balanced by a large number of smaller negative contributions in accordance with Eq.~(\ref{eq:DE_coh_sum}). The inset shows that, when negative, the coherent contribution (red) cannot exceed the incoherent one (blue) in absolute value, see Eq.~(\ref{eq:DE_coh_negative}). In both plots symbols / lines represent one disorder realization / statistically averaged results respectively.}
		\label{fig:coherent_contribution}
	\end{figure}
	
	The quantity $|d_{\beta\alpha}|^2$ represents the incoherent (i.e. a real positive intensity) contribution due to $\beta$-th transmission eigenchannel to the $\alpha$-th deposition eigenchannel. We obtain the sum rule~\cite{SM} 
	\begin{equation}
	\sum_{\beta=1}^N |d_{\beta\alpha}|^2 = 1 = \sum_{\alpha=1}^N |d_{\beta\alpha}|^2.
	\label{eq:DE_incoh_weight_sum}
	\end{equation}
	This is a non-trivial result, since $|d_{\beta\alpha}|^2$ cannot be interpreted as a weight coefficient due to presence of the interference term in Eq.~(\ref{eq:DE_decomposition_coh_incoh}). Furthermore, it leads to an important constraint on this very term. By summing both sides of that equation and using Eqs.~(\ref{eq:DE_summation_rule},\ref{eq:DE_incoh_weight_sum}), we get 
	\begin{equation}
	\sum_{\alpha=1}^N \left[
	\sum_{\beta \neq \beta^\prime} 
	d_{\beta\alpha} \, d^*_{\alpha\beta^\prime} 
	E_{\beta}^{(T)}(\vec{r}) \left[E_{\beta^\prime}^{(T)}(\vec{r})\right]^* 
	\right] \equiv 0.
	\label{eq:DE_coh_sum}
	\end{equation}
	As shown in Ref.~\cite{2022_Bender_Deposition_eigenchannels}, the interference contribution can be quite large, even dominant in some cases. The relationship in Eq.~(\ref{eq:DE_coh_sum}) states that the sum of interference contributions to all deposition eigenchannels is, in fact, zero. To illustrate this point, we plot~\cite{SM} in Fig.~\ref{fig:coherent_contribution} the coherent contributions at the target depth $z_D=L/2$ for all deposition eigenchannels. Each contribution is normalized by $(1/N)\sum_{n=1}^{N}\left\langle\left| E^{(0)}_n(z_D) \right|^2\right\rangle$, which represents the unoptimized intensity at the target depth. A small number of deposition eigenchannels have large positive contributions. In contrast, the number of small negative contributions is large, to ensure the sum is equal to 0 in Eq.~(\ref{eq:DE_coh_sum}). 
	
	Again we stress that the above relationships apply for every disorder configuration and do not require any statistical averaging. Furthermore, because the left hand side of  Eq.~(\ref{eq:DE_decomposition_coh_incoh}) is a positively defined quantity, we note that
	\begin{equation}
	\sum_{\beta=1}^N \left|E_{\beta}^{(T)}(\vec{r})\right|^2 |d_{\beta\alpha}|^2 
	\geq
	-\sum_{\beta \neq \beta^\prime}^N d_{\beta\alpha} \, d^*_{\alpha\beta^\prime} E_{\beta}^{(T)}(\vec{r}) \left[E_{\beta^\prime}^{(T)}(\vec{r})\right]^*.
	\label{eq:DE_coh_negative}
	\end{equation}
	It illustrates that when the interference contribution does become negative, i.e. for a low-deposition eigenhcannel, it cannot exceed in absolute value the incoherent contribution, i.e. it cannot become dominant, see the inset in Fig.~\ref{fig:coherent_contribution}. However, such restriction does not apply for the high-deposition eigenchannel with positive coherent contribution, which can and, in fact, does become dominant in a diffusive medium for the deposition depth $z_D < L/2$, c.f. Fig.~\ref{fig:coherent_contribution} and Ref.~\cite{2022_Bender_Deposition_eigenchannels}. We note that such larger positive interference contributions are related to the PDF of the deposition eigenvalues. Since the PDF $P(\zeta)$ has a long tail at large $\zeta$ in diffusive systems, the number of positive contributions is small. Consequently, the sum rule in Eq.~(\ref{eq:DE_coh_sum}) dictates that such contributions must be large in order to balance the numerous negative ones.
	
	\section*{Conclusions}\label{sec:conclusions}
	
	Targeted delivery of electromagnetic energy inside a random-scattering system has important applications in imaging, optogenetics, photothermal therapy, etc. The deposition eigenchannels accomplish the goal of delivering maximal or minimal amount of energy to a target region of arbitrary size, shape and depth. Little is known about the spatial structure of the deposition eignchannels, albeit some progress has been made in understanding the spatial distribution of transmission eigenchannels~\cite{2015_Genack_Eigenchannels_Inside,2016_Ojambati_Fundamental_Mode_Experiment,2017_Koirala_Inverse_Design}. On the other hand, the PDF of deposition eigenvalues has been predicted by the filtered random matrix theory~\cite{2022_Bender_Deposition_eigenchannels}. The sum rules, presented in this work, establish a connection between the spatial structure of the deposition eigenchannels and the eigenvalues. They represent the rigorous constraints for any disorder realization of 2D and 3D scattering systems in both waveguide and slab geometry. 
	
	\section*{Acknowledgements}\label{sec:acknowledgements}
	
	\section*{Funding}\label{sec:funding}
	We acknowledge Arthur Goetschy, Chia-Wei Hsu, Hasan Y{\i}lmaz for stimulating discussions. This work was supported by the National Science Foundation under Grant Nos. DMR-1905442, DMR-1905465.
	
	\section*{Disclosures}\label{sec:disclosures}
	The authors declare no conflicts of interest.

	\section*{Data availability}\label{sec:data}
	All data underlying the results is presented in this paper.\\
	
	See Supplement 1 for supporting content.
	
	\bibliography{2022_Deposition_eigenchannels_properties}

\begin{thebibliography}{10}
\newcommand{\enquote}[1]{``#1''}

\bibitem{2012_Mosk_SLM_review}
P.~A. Mosk, A.~Lagendijk, G.~Lerosey, and M.~Fink, {\protect\JournalTitle{Nat.
  Photon.}} \textbf{6}, 283 (2012).

\bibitem{2015_Yu_Wavefront_Shaping_Review}
H.~Yu, J.~Park, K.~Lee, J.~Yoon, K.~Kim, S.~Lee, and Y.~K. Park,
  {\protect\JournalTitle{Curr. Appl. Phys.}} \textbf{15}, 632 (2015).

\bibitem{kubby2019wavefront}
J.~Kubby, S.Gigan, and M.~Cui, \emph{Wavefront shaping for biomedical imaging}
  (Cambridge University Press, 2019).

\bibitem{2021_Gigan_Roadmap}
S.~Gigan, O.~Katz, H.~B. de~Aguiar, E.~R. Andresen, A.~Aubry, J.~Bertolotti,
  E.~Bossy, D.~Bouchet, J.~Brake, S.~Brasselet, Y.~Bromberg, H.~Cao,
  T.~Chaigne, Z.~Cheng, W.~Choi, T.~Čižmár, M.~Cui, V.~R. Curtis,
  H.~Defienne, M.~Hofer, R.~Horisaki, R.~Horstmeyer, N.~Ji, A.~K. LaViolette,
  J.~Mertz, C.~Moser, A.~P. Mosk, N.~C. Pégard, R.~Piestun, S.~Popoff, D.~B.
  Phillips, D.~Psaltis, B.~Rahmani, H.~Rigneault, S.~Rotter, L.~Tian, I.~M.
  Vellekoop, L.~Waller, L.~Wang, T.~Weber, S.~Xiao, C.~Xu, A.~Yamilov, C.~Yang,
  and H.~Yılmaz, {\protect\JournalTitle{arXiv}} 2111.14908 (2021).

\bibitem{2017_Rotter_Gigan_review}
S.~Rotter and S.~Gigan, {\protect\JournalTitle{Rev. Mod. Phys.}} \textbf{89},
  015005 (2017).

\bibitem{2015_Vellekoop_review}
I.~M. Vellekoop, {\protect\JournalTitle{Opt.~Express}} \textbf{23}, 12189
  (2015).

\bibitem{2010_Popoff_Shaping_PRL}
S.~M. Popoff, G.~Lerosey, R.~Carminati, M.~Fink, A.~C. Boccara, and S.~Gigan,
  {\protect\JournalTitle{Phys.~Rev.~Lett.}} \textbf{104}, 100601 (2010).

\bibitem{2015_Yang_Wavefront_Shaping_Guidestar}
R.~Horstmeyer, H.~Ruan, and C.~Yang, {\protect\JournalTitle{Nat.~Photon.}}
  \textbf{9}, 563 (2015).

\bibitem{yoon2020deep}
S.~Yoon, M.~Kim, M.~Jang, Y.~Choi, W.~Choi, S.~Kang, and W.~Choi,
  {\protect\JournalTitle{Nature Reviews Physics}} \textbf{2}, 141 (2020).

\bibitem{2014_Cheng_Wavefront_Shaping_Energy_Deposition}
X.~Cheng and A.~Z. Genack, {\protect\JournalTitle{Opt. Lett.}} \textbf{39},
  6324 (2014).

\bibitem{2022_Bender_Deposition_eigenchannels}
N.~Bender, A.~Yamilov, A.~Goetschy, H.~Yilmaz, C.~W. Hsu, and H.~Cao,
  {\protect\JournalTitle{Nat. Phys.}} \textbf{18}, 309 (2022).

\bibitem{SM}
Supplementary Material.

\bibitem{1984_Dorokhov}
O.~N. Dorokhov, {\protect\JournalTitle{Solid State Comm.}} \textbf{51}, 381
  (1984).

\bibitem{2015_Genack_Eigenchannels_Inside}
M.~Davy, Z.~Shi, J.~Park, C.~Tian, and A.~Z. Genack,
  {\protect\JournalTitle{Nat.~Comm.}} \textbf{6}, 6893 (2015).

\bibitem{2013_Stone_Eigenvalues_with_Absorption}
A.~Goetschy and A.~D. Stone, {\protect\JournalTitle{Phys. Rev. Lett.}}
  \textbf{111}, 063901 (2013).

\bibitem{2016_Ojambati_Fundamental_Mode_Experiment}
O.~S. Ojambati, H.~Yilmaz, A.~Lagendijk, A.~P. Mosk, and W.~L. Vos,
  {\protect\JournalTitle{New~J.~Phys.}} \textbf{18}, 043032 (2016).

\bibitem{2017_Koirala_Inverse_Design}
M.~Koirala, R.~Sarma, H.~Cao, and A.~Yamilov, {\protect\JournalTitle{Phys. Rev.
  B}} \textbf{96}, 054209 (2017).

\bibitem{2016_Sarma_Open_Channels}
R.~Sarma, A.~Yamilov, S.~Petrenko, Y.~Bromberg, and H.~Cao,
  {\protect\JournalTitle{Phys.~Rev.~Lett.}} \textbf{117}, 086803 (2016).

\bibitem{2020_Bender_correlations_PRL}
N.~Bender, A.~Yamilov, H.~Yilmaz, and H.~Cao, {\protect\JournalTitle{Phys. Rev.
  Lett.}} \textbf{125}, 165901 (2020).

\end{thebibliography}

	\newpage
	\onecolumn
	\setcounter{page}{1}
	\setcounter{equation}{0}
	\renewcommand{\theequation}{S\arabic{equation}}
	
	\section*{Supplement 1: Sum rules for energy deposition eigenchannels in scattering systems}\label{sec:SM}
    Alexey Yamilov$^1$, Nicholas Bender$^2$, Hui Cao$^3$\\
    $^1$Physics Department, Missouri University of Science \& Technology, Rolla, Missouri, USA\\
    $^2$School of Applied and Engineering Physics, Cornell University, Ithaca, New York, USA\\
    $^3$Department of Applied Physics, Yale University, New Haven, Connecticut, USA\\
	
	\subsection*{Derivation of Eq.~(\ref{eq:DE_summation_rule})}
	Taking advantage of the unitarity of $V^{(D)}$ matrices, $\sum_{n=1}^N \left[V_{\beta n}^{(D)}(\vec{r}_D)\right]^* V_{n\alpha}^{(D)}(\vec{r}_D)=\delta_{\alpha\beta}$, where $\delta_{\alpha\beta}$ is the Kronecker delta, we obtain
	\begin{equation}
	\begin{split}
	&\sum_{\alpha=1}^{N}\left| E_{\alpha}^{(D)}(\vec{r};\vec{r}_D) \right|^2\\
	&=\sum_{\alpha,nn^\prime=1}^{N} E^{(0)}_n(\vec{r}) \,  V_{n\alpha}^{(D)}(\vec{r}_D) \, 
	\left[V_{\alpha n^\prime}^{(D)}(\vec{r}_D)\right]^* \, \left[E^{(0)}_{n^\prime}(\vec{r})\right]^* \\
	&=\sum_{nn^\prime=1}^{N} E^{(0)}_n(\vec{r}) \,  \delta_{nn^\prime} \, \left[E^{(0)}_{n^\prime}(\vec{r})\right]^*=\sum_{n=1}^{N}\left| E^{(0)}_n(\vec{r}) \right|^2.
	\label{eq:sm:DE_summation_rule}
	\end{split}
	\end{equation}
	A similar relationship between intensities of TE $\left| E_\alpha^{(T)}(\vec{r}) \right|^2$ and $\left| E^{(0)}_n(\vec{r}) \right|^2$ can also be obtained analogously completing the derivation of Eq.~(\ref{eq:DE_summation_rule}).
	
	\subsection*{Derivation of Eq.~(\ref{eq:zeta_summation_rule})}
	Singular values $\zeta_\alpha(\vec{r}_D)$ can be extracted from $\mathcal{Z}^{\dagger}(\vec{r}_D)\mathcal{Z}(\vec{r}_D)$ matrix. Using the unitarity of $U^{(D)}(\vec{r}_D)$, $V^{(D)}(\vec{r}_D)$ matrices and the definition of $\mathcal{Z}(\vec{r}_D)$ in Eq.~(\ref{eq:Z_definition}), we can find an explicit relationship for the eigenvalue 
	\begin{equation}
	\begin{split}
	&\zeta_\alpha(\vec{r}_D)=
	\sum_{nn^\prime=1}^{N}\sum_{m=1}^{M} 
	\left[V_{\alpha n^\prime}^{(D)}(\vec{r}_D)\right]^*
	\mathcal{Z}_{n^\prime m}^{*}(\vec{r}_D)
	\mathcal{Z}_{mn}(\vec{r}_D)
	V_{n\alpha}^{(D)}(\vec{r}_D) \\
	&=
	\frac{\mathcal{V}}{M}\, \sum_{nn^\prime=1}^{N}\sum_{m=1}^{M} \epsilon(\vec{r}_m)
	\left[V_{\alpha n^\prime}^{(D)}(\vec{r}_D)\right]^*
	\left|E_n(\vec{r}_m;\vec{r}_D)\right|^2
	V_{n\alpha}^{(D)}(\vec{r}_D) \\
	&=
	\frac{\mathcal{V}}{M}\,\sum_{m=1}^{M} \epsilon(\vec{r}_m)
	\left|E_\alpha^{(D)}(\vec{r}_m;\vec{r}_D)\right|^2. 
	\end{split}\label{eq:sm:zeta_definition}
	\end{equation}
	We stress that this relationship is exact and holds in every disorder realization.
	
	\subsection*{Derivation of Eq.~(\ref{eq:DE_incoh_weight_sum})}
	Recalling the definition $d_{\beta\alpha}(\vec{r}_D) = \sum_{n^\prime=1}^{N}  \left[V_{\beta n^\prime}^{(T)}\right]^* V_{n^\prime\alpha}^{(D)}(\vec{r}_D)$, the quantity $|d_{\beta\alpha}|^2$ can be evaluated as
	\begin{equation}
	\begin{split}
	\sum_{\beta=1}^N |d_{\beta\alpha}|^2 
	&\equiv\sum_{nn^\prime\beta=1}^N V_{n\beta}^{(T)} \left[V_{\beta n^\prime}^{(T)}\right]^* V_{n^\prime\alpha}^{(D)}(\vec{r}_D) V_{\alpha n}^{(D)*}(\vec{r}_D)\\
	&\equiv\sum_{nn^\prime=1}^N \delta_{nn^\prime} V_{n^\prime\alpha}^{(D)}(\vec{r}_D) \left[V_{\alpha n}^{(D)}(\vec{r}_D)\right]\\
	&\equiv \sum_{n=1}^N  V_{n\alpha}^{(D)}(\vec{r}_D)\left[V_{\alpha n}^{(D)}(\vec{r}_D)\right]^*\equiv 1,
	\end{split}\label{eq:sm:DE_incoh_weight_sum}
	\end{equation}
	where we used the unitarity of the $V$ matrices. Summation over the other index $\alpha$ yieslds the same result.
	
	\subsection*{Numerical simulation in the waveguide geometry}\label{sec:numerical_simulations}
	
	The results presented in Figs.~\ref{fig:eigenvalues},\ref{fig:eigenchannels},\ref{fig:coherent_contribution} are obtained in two-dimensional waveguide geometry, see Refs.~\cite{2022_Bender_Deposition_eigenchannels,2017_Koirala_Inverse_Design,2016_Sarma_Open_Channels,2020_Bender_correlations_PRL} for the detailed description of the numerical model. The parameters of in the simulation are chosen to represent a diffusive transport with no loss. Specifically, system length is $L/\ell\simeq 15$, where $\ell$ is the transport mean free path. The width of the waveguide $W/L=0.3$, the number of modes in the waveguide $N=56$. The value of the dimensionless conductance $g\simeq 5.3$.
	
	Transmission eigenchannels are obtained using singular value decomposition of the transmission matrix $\mathcal{T}$. The deposition matrix $\mathcal{Z}_{mn}(\vec{r}_D)$ is defined using Eq.~(\ref{eq:Z_definition}) by sampling fields at all spatial points $E^{(0)}_n(y_m,z_D)$ at a specific depth $z_D$. $y$ and $z$ are the transverse and the longitudinal coordinates respectively. Subsequently, the deposition eigenchannels are obtained using singular value decomposition of the rectangular matrix $\mathcal{Z}_{mn}(z_D)$. The results presented in Figs.~\ref{fig:eigenchannels},\ref{fig:coherent_contribution} are obtained for the target region at the middle of the disordered region $z_D=L/2$.
	
	All numerical results in Figs.~\ref{fig:eigenvalues}-\ref{fig:coherent_contribution} are statistically averaged over an ensemble of $1000$ random disorder realizations.

\end{document}